\documentclass[aps,preprint,tightenlines]{revtex4}%
\usepackage{amsfonts}
\usepackage{amsmath}
\usepackage{amssymb}
\usepackage{graphicx}%
\providecommand{\U}[1]{\protect \rule{.1in}{.1in}}
\begin{document}
\title[Polarons in cold quantum gases]{Feynman path-integral treatment of the BEC-impurity polaron}
\author{J.\ Tempere$^{1,2}$, W. Casteels$^{1}$, M.K. Oberthaler$^{3}$, S. Knoop$^{3}$,
E. Timmermans$^{4}$, J.T. Devreese$^{1}$}
\affiliation{$^{1}$ TFVS, Universiteit Antwerpen, Groenenborgerlaan 171, B2020 Antwerpen,\ Belgium}
\affiliation{$^{2}$ Lyman Laboratory of Physics, Harvard University, Cambridge MA02138, USA.}
\affiliation{$^{3}$ Kirchhoff-Institut f\"{u}r Physik, Universit\"{a}t Heidelberg, Im
Neuenheimer Feld 227, D-69120 Heidelberg, Germany.}
\affiliation{$^{4}$ Theoretical Division (T-4), Los Alamos National Laboratory, Los Alamos,
NM 87545, USA.}

\begin{abstract}
The description of an impurity atom in a Bose-Einstein condensate can be cast
in the form of Fr\"{o}hlich's polaron Hamiltonian, where the Bogoliubov
excitations play the role of the phonons. An expression for the corresponding
polaronic coupling strength is derived, relating the coupling strength to the
scattering lengths, the trap size and the number of Bose condensed atoms. This
allows to identify several approaches to reach the strong-coupling limit for
the quantum gas polarons, whereas this limit was hitherto experimentally
inaccessible in solids. We apply Feynman's path-integral method to calculate
for all coupling strengths the polaronic shift in the free energy and the
increase in the effective mass. The effect of temperature on these quantities
is included in the description. We find similarities to the acoustic polaron
results and indications of a transition between free polarons and self-trapped
polarons. The prospects, based on the current theory, of investigating the
polaron physics with ultracold gases are discussed for lithium atoms in a
sodium condensate.

\end{abstract}
\date{September 8, 2009}
\maketitle

\section{Introduction}

Quantum gases are used as an excellent testbed for many-body theory, and are
particularly useful to investigate strong-coupling regimes or strongly
correlated regimes that have remained out of reach in the solid
state\cite{dalibardreview}. The current work focuses on the physics of an
impurity in a Bose-Einstein condensate (BEC), as this system opens a new and
promising avenue to investigate long-standing problems in polaron
theory\cite{polaronreview}. In 1952 Fr\"{o}hlich, inspired by Landau's concept
of the polaron, derived a Hamiltonian that describes a charge carrier
(electron, hole) interacting with its self-induced polarization, in an ionic
crystal or a polar semiconductor\cite{Landau,Frohlich}. The Fr\"{o}hlich
polaron Hamiltonian has resisted exact analytical diagonalization since 1952
and became a laboratory to test methods of quantum field
theory\cite{polaronreview}. Tomonaga's canonical transformation was applied to
study the weak coupling regime. Bogoliubov tackled the polaron strong coupling
limit with one of his canonical transformations. Feynman used his path
integral formalism (and an ad-hoc variational principle) to develop a superior
all coupling approximation for the polaron\cite{Feynman}. These studies were
addressing the self-energy, the effective mass and the mobility of
Fr\"{o}hlich polarons. Studies of the optical absorption of a Fr\"{o}hlich
polaron were initiated by Evrard \emph{et al}. \cite{KED} for strong coupling
and by Gurevich \emph{et al}. \cite{Gurevich} for the weak coupling limit.
Devreese \emph{et al.} \cite{Optipol} calculated the optical absorption of the
Fr\"{o}hlich polaron at all coupling using Feynman path integrals and the
response formalism introduced in Ref. \cite{FHIP}. The results of Ref.
\cite{Optipol} reveal the internal excitation structure of the Fr\"{o}hlich
polaron that is absent in Ref. \cite{FHIP}.

However the largest Fr\"{o}hlich polaron coupling constant established in any
solid by experiment is not large enough to reveal the rich internal excitation
structure of the optical absorption predicted in Ref. \cite{Optipol}. The
possibility to tune the coupling strength in the BEC-impurity system presents
a marvellous challenge and opportunity to experimentally reveal the internal
excitation structure and its related resonances and scattering states
\textquotedblleft contained\textquotedblright \ in the Fr\"{o}hlich
field-theoretical polaron Hamiltonian. Quantum gases could therefore reveal
important and subtle characteristics (eigenstates, resonances,...) of a
Hamiltonian devised for a solid, however, that cannot be realized in a solid.
The observation of the spectra of the BEC-impurity Hamiltonian at large
coupling would be all the more interesting in view of recent studies of the
spectrum of the Fr\"{o}hlich polaron Hamiltonian with Diagrammatic Quantum
Monte-Carlo numerical techniques \cite{MishchenkoPRL} and with a new strong
coupling model \cite{DeFilippisPRL96} that explore the validity of the
Franck-Condon principle with increasing coupling strength. Similarly, the
transition between quasi-free polarons and self-trapped polarons predicted for
acoustic polarons\cite{acoustipol} is expected to occur in a coupling regime
hither\-to inaccessible in the solid state. Finally, a better understanding of
the intermediate and strong-coupling regimes is needed to further elucidate
the role of polarons and bipolarons in unconventional pairing mechanisms for
high-temperature superconductivity\cite{alexandrov}.

The Hamiltonian of an impurity in a Bose-Einstein condensate can be mapped
onto the polaron Hamiltonian when the Bogoliubov approximation is
valid\cite{TimmermansPRL,TimmermansPRA}. The polaronic effects comes about
through the coupling of the impurity with the Bogoliubov excitations, as shown
in Sec. II. We apply the path-integral formalism, valid at all values of the
coupling strength, to describe the polaronic effect in a condensate in Sec.
III. As a concrete example, we use the parameters for a lithium impurity in a
sodium condensate, and derive an expression linking the experimental
parameters to a dimensionless coupling constant analogous to the Fr\"{o}hlich
coupling constant for electrons in polar crystals. The polaronic energy shift,
the effective mass increase and the polaron radius are studied for all values
of this coupling strength and at all temperatures where the Bogoliubov
approximation is valid, in Sec. IV.

\section{The polaron Hamiltonian for an impurity in a condensate}

The Hamiltonian of a single atom, in the presence of a Bose gas is given by:
\begin{equation}
\hat{H}=\frac{\hat{p}^{2}}{2m_{I}}+\sum_{\mathbf{k}}\varepsilon_{\mathbf{k}%
}\hat{a}_{\mathbf{k}}^{\dag}\hat{a}_{\mathbf{k}}+\frac{1}{2}\sum
_{\mathbf{k},\mathbf{k}^{\prime}\mathbf{,q}}V_{BB}(\mathbf{q})\hat
{a}_{\mathbf{k}^{\prime}-\mathbf{q}}^{\dag}\hat{a}_{\mathbf{k+q}}^{\dag}%
\hat{a}_{\mathbf{k}}\hat{a}_{\mathbf{k}^{\prime}}+\sum_{\mathbf{k},\mathbf{q}%
}V_{IB}(\mathbf{q})\hat{\rho}_{I}(\mathbf{q})\hat{a}_{\mathbf{k}^{\prime
}\mathbf{-q}}^{\dag}\hat{a}_{\mathbf{k}^{\prime}} \label{Hgas}%
\end{equation}
The first term represents the kinetic energy of the "impurity" atom with mass
$m_{I}$. The operators $\hat{a}_{\mathbf{k}}^{\dag},\hat{a}_{\mathbf{k}}$
create and annihilate a boson with mass $m_{B}$, wave number \textbf{k}, and
energy $\varepsilon_{\mathbf{k}}=(\hbar k)^{2}/(2m_{B})-\mu$ where $\mu$ is
the chemical potential. These bosons interact, and $V_{BB}(\mathbf{q})$ is the
Fourier transform of the boson-boson interaction potential. The interaction
between the bosonic atoms and the impurity atom is described by a potential
$V_{IB}(\mathbf{q})$ coupling the boson density to the impurity density
$\hat{\rho}_{I}(q)$ which can be expressed as a function of the impurity
position operator $\hat{r}$ as%
\begin{equation}
\hat{\rho}_{I}(\mathbf{q})=\int d^{3}\mathbf{r}\text{ }e^{-i\mathbf{q\cdot r}%
}\delta(\mathbf{r}-\mathbf{\hat{r}}).
\end{equation}

Bose-Einstein condensation is realized when the single-particle density matrix
has an eigenvalue $N_{0}\gg1$ comparable to the total number of bosons
\cite{PenroseOnsager}. For a homogeneous condensate, this is expressed by the
Bogoliubov shift \cite{Bogoliubov} which transforms the Hamiltonian
(\ref{Hgas}) into%
\begin{equation}
\hat{H}=E_{GP}+N_{0}V_{IB}(\mathbf{0})+\dfrac{\hat{p}^{2}}{2m_{I}}+%
%TCIMACRO{\dsum \limits_{\mathbf{k}\neq0}}%
%BeginExpansion
{\displaystyle \sum \limits_{\mathbf{k}\neq0}}
%EndExpansion
E_{\mathbf{k}}\hat{b}_{\mathbf{k}}^{\dag}\hat{b}_{\mathbf{k}}+%
%TCIMACRO{\dsum \limits_{\mathbf{k}\neq0}}%
%BeginExpansion
{\displaystyle \sum \limits_{\mathbf{k}\neq0}}
%EndExpansion
\sqrt{\dfrac{\xi_{\mathbf{k}}N_{0}}{E_{\mathbf{k}}}}V_{IB}(\mathbf{k}%
)\hat{\rho}_{I}(\mathbf{k})\left[  \hat{b}_{\mathbf{k}}+\hat{b}_{-\mathbf{k}%
}^{\dag}\right]  \label{Hbol}%
\end{equation}
Here, the operators $\hat{b}_{\mathbf{k}}^{\dag},\hat{b}_{\mathbf{k}}$ create
resp. annihilate Bogoliubov excitations with wave number $\mathbf{k}$ and
dispersion%
\begin{equation}
E_{\mathbf{k}}=\sqrt{\xi_{\mathbf{k}}\left[  \xi_{\mathbf{k}}+2N_{0}%
V_{BB}(\mathbf{k})\right]  }%
\end{equation}
where $\xi_{\mathbf{k}}=(\hbar k)^{2}/(2m_{B})$. The first term in
(\ref{Hbol}) represents the Gross-Pitaevskii energy \cite{Stringaribook}%
\begin{equation}
E_{GP}=N_{0}\varepsilon_{0}+\frac{N_{0}^{2}}{2}V_{BB}(\mathbf{0})+\frac{1}{2}%
%TCIMACRO{\dsum \limits_{\mathbf{k}\neq0}}%
%BeginExpansion
{\displaystyle \sum \limits_{\mathbf{k}\neq0}}
%EndExpansion
N_{0}V_{BB}(\mathbf{k}) \label{add}%
\end{equation}
and the second term is the interaction shift due to the impurity. For both the
boson-boson interaction and the boson-impurity interaction we will assume
contact pseudopotentials: $V_{BB}(\mathbf{r}-\mathbf{r}^{\prime})=g_{BB}%
\delta(\mathbf{r}-\mathbf{r}^{\prime})$ and $V_{IB}(\mathbf{r}-\mathbf{r}%
^{\prime})=g_{IB}\delta(\mathbf{r}-\mathbf{r}^{\prime})$. The interaction
strengths $g_{BB}$ and $g_{IB}$ are related to the boson-boson scattering
length $a_{BB}$ and the boson-impurity scattering length $a_{IB},$
respectively, through the Lippmann-Schwinger equation. For the boson-boson
interaction, the first order result $g_{BB}=4\pi \hbar^{2}a_{BB}/m_{B}$ will
suffice. For the boson-impurity interaction, the Lippmann-Schwinger equation
needs to be treated correctly up to second order to obtain valid results for
the polaron problem, since as we shall see, $g_{IB}^{2}$ will appear in the expressions.

The resulting Hamiltonian (\ref{Hbol}) maps onto the Fr\"{o}hlich polaron
Hamiltonian \cite{Frohlich}%
\begin{equation}
\hat{H}_{pol}=\dfrac{\hat{p}^{2}}{2m_{I}}+%
%TCIMACRO{\dsum \limits_{\mathbf{k}\neq0}}%
%BeginExpansion
{\displaystyle \sum \limits_{\mathbf{k}\neq0}}
%EndExpansion
\hbar \omega_{\mathbf{k}}\hat{b}_{\mathbf{k}}^{\dag}\hat{b}_{\mathbf{k}}+%
%TCIMACRO{\dsum \limits_{\mathbf{k}\neq0}}%
%BeginExpansion
{\displaystyle \sum \limits_{\mathbf{k}\neq0}}
%EndExpansion
V_{\mathbf{k}}e^{i\mathbf{k\cdot \hat{r}}}\left(  \hat{b}_{\mathbf{k}}+\hat
{b}_{-\mathbf{k}}^{\dag}\right)  \label{Hpol}%
\end{equation}
with
\begin{align}
\hbar \omega_{\mathbf{k}}  &  =ck\sqrt{1+(\xi k)^{2}/2}\label{wk}\\
V_{\mathbf{k}}  &  =\sqrt{N_{0}}\left[  \dfrac{(\xi k)^{2}}{(\xi k)^{2}%
+2}\right]  ^{1/4}g_{IB} \label{Vk}%
\end{align}
In these expressions $\xi=1/\sqrt{8\pi a_{BB}n_{0}}$ is the healing length of
the Bose condensate with $n_{0}=N_{0}/V$ the condensate density (in the
present calculatations for the homogeneous gas we work with unit volume) and
$c=\hbar/(\sqrt{2}m_{B}\xi)$ is the speed of sound in the condensate. The
operator structure of the BEC-impurity Hamiltonian and that of the
Fr\"{o}hlich polaron Hamiltonian is identical. Depending on the analytical
form of the scalar functions multiplying the field operators, the Fr\"{o}hlich
polaron Hamiltonian, originally devised to describe the electron/hole -
longitudinal optical phonon interaction, can depict as well the
acousto-polaron\cite{acoustic1,acoustipol}, the piezo-polaron\cite{Mahan}, the
ripplo-polaron \cite{ripplopol}, etc... However, a formalism/method that
diagonalizes the Fr\"{o}hlich Hamiltonian at all coupling, does so for all
those different types of polarons, including the BEC-impurity polaron.

This mapping of the BEC-impurity problem onto the Fr\"{o}hlich Hamiltonian,
and the resulting possibility of polaronic self-trapping of an impurity in a
condensate, has been the subject of recent theoretical investigations at
strong coupling \cite{TimmermansPRL,TimmermansPRA}, and at low coupling with
the Lee-Low-Pines scheme \cite{Huang}. For small polarons in an optical
lattice, a perturbative study exists \cite{BrudererPRA}. As is also known from
the study of the polaronic problem for slow electrons in a polar crystal,
these approximations give qualitatively different results for different
regimes of coupling strength. These different characteristics between
weak-coupling and strong-coupling behavior are even more pronounced in the
case of an impurity in a condensate. In this work, we present a calculation
valid for all coupling strengths, based on the Feynman-Jensen variational
scheme first successfully applied by Feynman \cite{Feynman} to study slow
electrons interacting with optical phonons in polar crystals. As discussed
below, the present system will have more similarities to acoustic polarons
than to the Fr\"{o}hlich polarons.

\bigskip

\section{Jensen-Feynman free energy}

\subsection{General treatment}

The calculation of the free energy $F$ of the polaron is based on the
Feynman-Jensen variational inequality \cite{Feynmanbook,Kleinertbook}%
\begin{equation}
F\leqslant F_{0}+\frac{1}{\hbar \beta}\left \langle \mathcal{S}-\mathcal{S}%
_{0}\right \rangle _{\mathcal{S}_{0}} \label{JensFeyn}%
\end{equation}
where $\mathcal{S}$ is the action functional of the real system described by
the Hamiltonian (\ref{Hpol}), and $\mathcal{S}_{0}$ is the action functional
of a variational model system with free energy $F_{0}$, and $\beta=1/(k_{B}T)$
with $T$ the temperature. This variational principle is an extension of the
usual Gibbs-Bogoliubov variational principle $F\leqslant F_{0}+\left \langle
\hat{H}-\hat{H}_{0}\right \rangle $ where $\hat{H}$ is the Hamiltonian of the
system under study and $\hat{H}_{0}$ is a model Hamiltonian, and the brackets
indicate statistical averaging. The inequality \ref{JensFeyn} can be
derived\cite{Feynmanbook,Kleinertbook} from the path-integral expression for
the partition sum:
\begin{align}
e^{-\beta F}  &  =\int \mathcal{D}\mathbf{r}\text{ }e^{-\mathcal{S}%
[\mathbf{r}(t)]/\hbar}\nonumber \\
&  =\int \mathcal{D}\mathbf{r}\text{ }e^{-\left \{  \mathcal{S}[\mathbf{r}%
(t)]-\mathcal{S}_{0}[\mathbf{r}(t)]\right \}  /\hbar}e^{-\mathcal{S}%
_{0}[\mathbf{r}(t)]/\hbar}\nonumber \\
&  =e^{-\beta F_{0}}\left \langle e^{-\left(  \mathcal{S}-\mathcal{S}%
_{0}\right)  /\hbar}\right \rangle _{\mathcal{S}_{0}}\nonumber \\
&  \geqslant \exp \left[  -\beta F_{0}-\left \langle \mathcal{S}-\mathcal{S}%
_{0}\right \rangle _{\mathcal{S}_{0}}/\hbar \right]
\end{align}
where the inequality follows from the convexity of the exponential, and
$\mathcal{D}\mathbf{r}$ represents the path-integral measure. The peculiarity
of this variational principle, in comparison to the usual Rayleigh-Ritz
variational approach, is that a variational action functional is used rather
than a variational wave function. One important strength of the Feynman-Jensen
principle is that it can be used at nonzero temperatures\cite{Feynmanbook}.
Another strength, in comparison with the Gibbs-Bogoliubov variational
principle, is that it can treat systems that are hard or impossible to
describe in the standard Hamiltonian formalism, such as a collection of
particles that interact through a retarded potential, such as Lienard-Wiechert
potentials in electrodynamics, or the phonon-induced electron-electron
potential in polaron theory. In particular, the polaron Hamiltonian
(\ref{Hpol}) after the elimination of the phonon degrees of freedom leads to
an action functional $\mathcal{S}$ containing retardation effects:%
\begin{equation}
\mathcal{S}=%
%TCIMACRO{\dint \limits_{0}^{\hbar\beta}}%
%BeginExpansion
{\displaystyle \int \limits_{0}^{\hbar \beta}}
%EndExpansion
\dfrac{m_{I}}{2}\dot{r}^{2}(\tau)d\tau-%
%TCIMACRO{\dsum \limits_{\mathbf{k}\neq0}}%
%BeginExpansion
{\displaystyle \sum \limits_{\mathbf{k}\neq0}}
%EndExpansion
\dfrac{\left \vert V_{\mathbf{k}}\right \vert ^{2}}{2\hbar}%
%TCIMACRO{\dint \limits_{0}^{\hbar\beta}}%
%BeginExpansion
{\displaystyle \int \limits_{0}^{\hbar \beta}}
%EndExpansion
d\tau%
%TCIMACRO{\dint \limits_{0}^{\hbar\beta}}%
%BeginExpansion
{\displaystyle \int \limits_{0}^{\hbar \beta}}
%EndExpansion
d\sigma \text{ }\mathcal{G}(\mathbf{k},|\tau-\sigma|)e^{i\mathbf{k\cdot}\left[
\mathbf{r}(\tau)-\mathbf{r}(\sigma)\right]  }. \label{realsys}%
\end{equation}
with $\mathcal{G}(\mathbf{k},u)$ the Bogoliubov excitation Green's function,
given by
\begin{equation}
\mathcal{G}(\mathbf{k},u)=\frac{\cosh \left[  \omega_{\mathbf{k}}(u-\hbar
\beta/2)\right]  }{\sinh(\hbar \beta \omega_{\mathbf{k}}/2)},
\end{equation}
The action functional (\ref{realsys}) is obtained by integrating out
analytically the oscillator degrees of freedom corresponding to the Bogoliubov
excitations, a technique introduced in the context of phonons in Ref.
\cite{Feynman}. This gives rise to a retarded interaction (proportional to
$e^{i\mathbf{k\cdot}\left[  \mathbf{r}(\tau)-\mathbf{r}(\sigma)\right]  }$),
mediated by the Bogoliubov-excitation Green's function, and with coupling
strength $\left \vert V_{\mathbf{k}}\right \vert ^{2}/(2\hbar)$.

The system under study here is modelled by a variational trial system with
free energy $F_{0}$ described by the action functional
\begin{equation}
\mathcal{S}_{0}=%
%TCIMACRO{\dint \limits_{0}^{\hbar\beta}}%
%BeginExpansion
{\displaystyle \int \limits_{0}^{\hbar \beta}}
%EndExpansion
\dfrac{m_{I}}{2}\dot{r}^{2}(\tau)d\tau+\dfrac{MW^{3}}{8}%
%TCIMACRO{\dint \limits_{0}^{\hbar\beta}}%
%BeginExpansion
{\displaystyle \int \limits_{0}^{\hbar \beta}}
%EndExpansion
d\tau%
%TCIMACRO{\dint \limits_{0}^{\hbar\beta}}%
%BeginExpansion
{\displaystyle \int \limits_{0}^{\hbar \beta}}
%EndExpansion
d\sigma \dfrac{\cosh \left[  W\left(  |\tau-\sigma|-\hbar \beta/2\right)
\right]  }{\sinh \left(  \beta \hbar W/2\right)  }\left[  \mathbf{r}%
(\tau)-\mathbf{r}(\sigma)\right]  ^{2}. \label{model}%
\end{equation}
The model system corresponds to a mass $m_{I}$ that is coupled by a spring
with spring constant $MW^{2}$ to a "Bogoliubov" mass $M.$ Both $M$ and $W$ are
variational parameters. The expectation value in the inequality is evaluated
in the model system. This model system is chosen partly because the
expectation values can be calculated analytically for it. For more details on
the formalism, we refer to Ref. \cite{Feynman,Feynmanbook,Schultz}:
\begin{align}
F  &  \leqslant \frac{3}{\beta}\left \{  \ln \left[  2\sinh \left(  \frac
{\beta \hbar \Omega}{2}\right)  \right]  -\ln \left[  2\sinh \left(  \frac
{\beta \hbar W}{2}\right)  \right]  \right \} \nonumber \\
&  -\frac{3}{2\beta}\frac{M}{m_{I}+M}\left[  \frac{\hbar \beta \Omega}{2}%
\coth \left(  \frac{\beta \hbar \Omega}{2}\right)  -1\right] \nonumber \\
&  -\sum_{\mathbf{k}}\frac{\left \vert V_{\mathbf{k}}\right \vert ^{2}}{\hbar}%
%TCIMACRO{\dint \limits_{0}^{\hbar\beta}}%
%BeginExpansion
{\displaystyle \int \limits_{0}^{\hbar \beta}}
%EndExpansion
du\left(  1-\frac{u}{\hbar \beta}\right)  \mathcal{G}(\mathbf{k},u)\mathcal{M}%
_{M,\Omega}(\mathbf{k},u). \label{FreeF}%
\end{align}
Here $\Omega=W\sqrt{1+M/m_{I}}$, and the memory function is given by%
\begin{equation}
\mathcal{M}_{M,\Omega}(\mathbf{k},u)=\exp \left[  -\frac{\hbar k^{2}}%
{2(m_{I}+M)}\left(  u-\frac{u^{2}}{\hbar \beta}-\frac{M}{m_{I}}\frac
{\cosh \left[  \Omega(\hbar \beta/2)\right]  -\cosh \left[  \Omega(\hbar
\beta/2-u)\right]  }{\Omega \sinh(\beta \hbar \Omega/2)}\right)  \right]  .
\end{equation}
Rather than choosing $W,M$ as variational parameters, we can choose $\Omega,M$
as variational parameters.

\subsection{BEC-impurity}

Care should be taken when we substitute\ the interaction amplitude (\ref{Vk})
into these expressions. Since $\left \vert V_{\mathbf{k}}\right \vert
^{2}\propto g_{IB}^{2},$ we need to solve the Lippmann-Schwinger equation up
to \emph{second} order to obtain the link between $g_{IB}$ and $a_{IB}$
correctly; we find%
\begin{equation}
\frac{2\pi \hbar^{2}a_{IB}}{m_{r}}=g_{IB}-g_{IB}^{2}\sum_{\mathbf{k}^{\prime
}\neq0}\frac{2m_{r}}{(\hbar k^{\prime})^{2}},
\end{equation}
where $m_{r}$ is the relative mass ($m_{r}^{-1}=m_{B}^{-1}+m_{I}^{-1}$). From
this, we find that the term $N_{0}V_{IB}(\mathbf{0})=N_{0}g_{IB}$ in
(\ref{Hbol}) must also contribute a renormalization factor since%
\begin{equation}
N_{0}g_{IB}\rightarrow N_{0}\left(  \frac{2\pi \hbar^{2}a_{IB}}{m_{r}}%
+g_{IB}^{2}\sum_{\mathbf{k}^{\prime}\neq0}\frac{2m_{r}}{(\hbar k^{\prime}%
)^{2}}\right)  . \label{reno}%
\end{equation}
Substituting (\ref{Vk}) and (\ref{wk}) into (\ref{FreeF}) then yields%
\begin{align}
&  F\leq \frac{3}{\beta}\left \{  \ln \left[  2\sinh \left(  \frac{\beta \Omega}%
{2}\right)  \right]  -\ln \left[  2\sinh \left(  \frac{\beta \Omega}{2\sqrt{1+M}%
}\right)  \right]  \right \} \label{result}\\
&  -\frac{3}{2\beta}\frac{M}{1+M}\left[  \frac{\beta \Omega}{2}\coth \left(
\frac{\beta \Omega}{2}\right)  -1\right] \nonumber \\
&  +\alpha \left(  \frac{\tilde{m}_{B}+1}{\tilde{m}_{B}}\right)  ^{2}%
%TCIMACRO{\dint \limits_{0}^{\infty}}%
%BeginExpansion
{\displaystyle \int \limits_{0}^{\infty}}
%EndExpansion
\frac{dk}{4\pi}\left[  \left(  \frac{2\tilde{m}_{B}}{1+\tilde{m}_{B}}\right)
-\frac{k^{3}}{\sqrt{2+k^{2}}}%
%TCIMACRO{\dint \limits_{0}^{\beta/2}}%
%BeginExpansion
{\displaystyle \int \limits_{0}^{\beta/2}}
%EndExpansion
\mathcal{G}(k,u)\mathcal{M}(k,u)du\right]  .\nonumber
\end{align}
In this expression,
\begin{equation}
\alpha=a_{IB}^{2}/(a_{BB}\xi) \label{coupl}%
\end{equation}
is the polaronic coupling strength. We use polaronic units, so that
$\hbar=m_{I}=\xi=1$. That means the free energy is in units $\hbar^{2}%
/(m_{I}\xi^{2})$, so that $\beta=\hbar^{2}/(m_{I}\xi^{2}k_{B}T)$ and
$\tilde{m}_{B}=m_{B}/m_{I}$. The integration variables $k,u$ and the
variational parameters $M,\Omega$ in (\ref{result}) are dimensionless. In
these units, we find after substitution of (\ref{Vk}) and (\ref{wk}), that the
Green's function is given by%
\[
\mathcal{G}(k,u)=\dfrac{\cosh \left[  k\sqrt{k^{2}+2}(2u-\beta)/\left(
4\tilde{m}_{B}\right)  \right]  }{\sinh \left[  \beta k\sqrt{k^{2}+2}/\left(
4\tilde{m}_{B}\right)  \right]  },
\]
and the memory function takes the form
\begin{equation}
\mathcal{M}(k,u)=\exp \left[  -\frac{k^{2}}{2(1+M)}\left(  \frac{u(\beta
-u)}{\beta}+M\frac{\cosh \left[  \Omega(\beta/2)\right]  -\cosh \left[
\Omega(\beta/2-u)\right]  }{\Omega \sinh(\beta \Omega/2)}\right)  \right]  .
\end{equation}
The first term of the integrand of the $k$-integral in (\ref{result}) is due
to a renormalization factor which arises when we relate $g_{IB}$ to the
boson-impurity scattering length. This term is independent of the variational
parameters $M$ and $\Omega$ and does not influence the optimization of the
variational parameters; it is however necessary for the $k$-integrals to
converge. In the case of the acoustic polaron, convergence in the variational
minimization of (\ref{result}) is sped up by introducing a cut-off $K_{c}$ to
the wave-number integral\cite{acoustic1,acoustipol}, for which there is a
natural choice, namely the edge of the Brillouin zone. For atomic gases, as we
show in Sec. IV, the results do not strongly depend on the value cut-off as
long as it is on the order of the (inverse) Van der Waals radius which we use
as a natural scale for $K_{c}$ in the present case. Note that, although these
formulae are derived for any temperature (any $\beta$), the mapping onto the
polaron Hamiltonian is only valid for temperatures below the critical
temperature, and low enough so that the number of Bogoliubov excitations can
be neglected with respect to the number of atoms in the condensate, so that
(\ref{Hbol}) gives a good description of the quantum gas. In the discussion,
we will look at realistic experimental values for $\beta$ and other parameters.

\subsection{Weak and strong coupling limits}

From our central result (\ref{result}), we can retrieve both the weak-coupling
and the strong-coupling results by a judicious choice of variational
parameters, or rather variational trial actions. The weak-coupling limit is
obtained by taking the limit $M\rightarrow0$ in the Jensen-Feynman treatment,
which corresponds to a the trial action
\begin{equation}
\mathcal{S}_{0}^{weak}=%
%TCIMACRO{\dint \limits_{0}^{\hbar\beta}}%
%BeginExpansion
{\displaystyle \int \limits_{0}^{\hbar \beta}}
%EndExpansion
\dfrac{m_{I}}{2}\dot{r}^{2}(\tau)d\tau
\end{equation}
of a free particle, as pointed out in the original work by Feynman
\cite{Feynman}. In the limit of zero temperature, this corresponds exactly to
the second-order perturbation result $E^{weak}(\alpha)$; we find
\begin{equation}
E^{weak}=\alpha \dfrac{(1+\tilde{m}_{B})^{2}}{\tilde{m}_{B}}\frac{1}{2\pi}%
%TCIMACRO{\dint \limits_{0}^{\infty}}%
%BeginExpansion
{\displaystyle \int \limits_{0}^{\infty}}
%EndExpansion
dq\left[  \frac{1}{1+\tilde{m}_{B}}-\sqrt{\frac{q^{2}}{q^{2}+2}}\frac
{q}{\tilde{m}_{B}q+\sqrt{q^{2}+2}}\right]  \label{weak}%
\end{equation}
Note that the variational energy becomes independent of $\Omega$ in the limit
$M\rightarrow0$. The energy grows linearly with $\alpha$, and the
proportionality constant depends on $m_{B}$ :%
\begin{equation}
E_{pol}^{weak}(\alpha)=\left \{
\begin{array}
[c]{c}%
\alpha/(\sqrt{2}\pi)\text{ for }m_{B}\rightarrow \infty \\
4\sqrt{2}\alpha/(3\pi)\text{ for }m_{B}\rightarrow1\\
\alpha/(\sqrt{2}m_{B})\text{ for }m_{B}\rightarrow0
\end{array}
\right.
\end{equation}
The strong-coupling limit is derived by taking the limit $M\rightarrow \infty$
in the Jensen-Feynman treatment. In this limit, the variational particle with
mass $M\rightarrow \infty$ is fixed in place, and we get a particle attached by
a spring to a fixed point rather than to a mobile variational mass. The
resulting trial action for strong coupling is\cite{Feynman}
\[
\mathcal{S}_{0}^{strong}=%
%TCIMACRO{\dint \limits_{0}^{\hbar\beta}}%
%BeginExpansion
{\displaystyle \int \limits_{0}^{\hbar \beta}}
%EndExpansion
\left(  \dfrac{m_{I}}{2}\dot{r}^{2}(\tau)+\dfrac{m_{I}\Omega^{2}}{2}r^{2}%
(\tau)\right)  d\tau.
\]
relating the strong-coupling limit to the variational approach of Landau and
Pekar\cite{Pekar1946} for the polaron problem, based on a Gaussian variational
wave function. The variational energy becomes%
\begin{equation}
E^{strong}\leq \frac{3}{4}\Omega+\lim_{\beta \rightarrow \infty}\alpha \left(
\frac{\tilde{m}_{B}+1}{\tilde{m}_{B}}\right)  ^{2}%
%TCIMACRO{\dint \limits_{0}^{\infty}}%
%BeginExpansion
{\displaystyle \int \limits_{0}^{\infty}}
%EndExpansion
\frac{dk}{4\pi}\left[  \left(  \frac{2\tilde{m}_{B}}{1+\tilde{m}_{B}}\right)
-\frac{k^{3}}{\sqrt{2+k^{2}}}%
%TCIMACRO{\dint \limits_{0}^{\beta/2}}%
%BeginExpansion
{\displaystyle \int \limits_{0}^{\beta/2}}
%EndExpansion
\mathcal{G}(k,u)\mathcal{M}(k,u)du\right]  \label{strong}%
\end{equation}
Care must be taken with the limit $\beta \rightarrow \infty$ : since the
convergence of $\mathcal{G}(k,u)$ to its $\beta \rightarrow \infty$ limit is not
uniform the limit can not be interchanged with the integration.

\bigskip

\section{Results and discussion}

\subsection{Experimental realization}

Before we turn to the results, it is useful to find typical experimental
values of the system parameters. Consider a sodium condensate of $N_{0}%
=10^{5}$ atoms in a harmonic trap with axial and radial trapping frequencies
$\omega_{\text{ax}}=2\pi \cdot50$ Hz and $\omega_{\text{rad}}=2\pi \cdot150$ Hz,
respectively. As scattering length for sodium in the $\left \vert
F,M_{F}\right \rangle =\left \vert 1,1\right \rangle $ hyperfine state, we use
$a_{BB}=2.8$ nm\cite{Samuelis}. The condensate is suitably described by the
Thomas-Fermi (TF) approximation. In our calculations we have assumed spatial
homogeneity. The density appears in the theoretical results through the
healing length, which is the relevant length-scale for the polaronic effect,
and appears in expression (\ref{coupl}) for the coupling constant. As long as
the polaronic effects are restricted to a range considerably smaller than the
TF radius (9.5 $\mu$m), we can use the local density at the center of the
trap, $n_{0}=7\times10^{13}$ cm$^{-3}$, to estimate the appropriate healing
length. In the current example, it is $\xi=450$ nm at the center of the trap,
indeed smaller than the TF radius. Using $\xi=1/\sqrt{8\pi n_{0}a_{BB}}$ and
the Thomas-Fermi expression for $n_{0}$ in the expression (\ref{coupl}), we
find the useful relation for the coupling constant
\begin{equation}
\alpha=15^{1/5}\left(  \frac{N_{0}a_{BB}}{\bar{a}_{HO}}\right)  ^{1/5}%
\frac{a_{IB}^{2}}{a_{BB}\bar{a}_{HO}} \label{alpha}%
\end{equation}
expressing the polaronic coupling constant as a function of the number of
condensate atoms $N_{0}$, the oscillator length $\bar{a}_{HO}=\sqrt
{\hbar/(m\omega_{HO})}$, associated with the geometrical average trap
frequency $\omega_{HO}=\sqrt[3]{\omega_{\text{ax}}\omega_{\text{rad}}^{2}}$,
and the $s$-wave scattering lengths $a_{BB}$ and $a_{IB}$.

For the impurity, we consider a $^{6}$Li atom, so that%
\begin{equation}
\tilde{m}_{B}=m_{B}/m_{I}=3.8
\end{equation}
Using $a_{IB}=0.8$ nm as the $^{6}$Li-$^{23}$Na scattering length
\cite{Gacesa}, $\alpha$ is of the order of $10^{-3}$. For these values, the
weak-coupling expression (\ref{weak}) is suitable. To increase $\alpha$,
several strategies are possible. One possibility is to increase $a_{IB}%
/a_{BB}$ using Feshbach resonances to either increase $a_{IB}$ or decrease
$a_{BB}$. One could also increase $a_{IB}/\bar{a}_{HO}$ by tightening the
trap. The remaining factor in expression (\ref{alpha}) varies slowly as a
function of the TF parameter $N_{0}a_{BB}/\bar{a}_{HO}$.

We look in particular at two strategies for increasing $\alpha$ in the case of
lithium impurities in a sodium condensate. On the one hand, one can minimize
$a_{BB}$ using the zero-crossing near the sodium Feshbach resonance at 907
G\cite{Inouye}. The tunability of $a_{BB}$ is limited by the magnetic field
stability; we expect a factor of 10 increase in $\alpha$ to be realistic. On
the other hand, one can increase $a_{IB}$ using a sodium-lithium Feshbach
resonance, for which the prime candidates are the already observed resonance
at 796 G \cite{Stan} or the predicted resonance at 1186 G \cite{Gacesa}. In
this strategy, the limiting factor to the tunability of $a_{IB}$ is the
trapping lifetime of the lithium atoms. The main loss process is three-body
recombination, involving one lithium and two sodium atoms, which is strongly
enhanced near the lithium-sodium resonance as the loss scales as $a_{IB}^{4}$.
We expect that an increase in $\alpha$ ($\propto a_{IB}^{2}$) up to a factor
1000 are still attainable. This leads to a maximal $\alpha$ of order unity,
which is in the crossover to the strong-coupling regime. In addition, to avoid
phase separation or collapse the trapping potential and the number of lithium
atoms have to be carefully chosen.

The energy scale $\hbar^{2}/(m_{I}\xi^{2})$ corresponds to a frequency of
$2\pi \cdot8.1$ kHz or a temperature of ca. $390$ nK. With this choice,
\begin{equation}
\beta=\frac{390\text{ nK}}{T\text{ (nK)}}%
\end{equation}
The (ideal gas) critical temperature for the sodium condensate is $220$ nK, or
$\beta=1.8$. As the lowest temperature we estimate $0.5$ $T_{\mu}$ where
$T_{\mu}=\mu/k_{B}=53$ nK. This corresponds to $\beta=15,$ so that the
experimentally relevant window for this parameter is $\beta \in \lbrack1.8,15]$.

The Bogoliubov dispersion at low $k$ is similar to that of acoustic phonons
rather than to that of optical phonons. For the acoustic polaron, it is known
that the variational parameters -especially the polaron mass- depend on the
value of a cut-off in the $K_{c}$ integral\cite{acoustipol}. For phonons in
solids, this cut-off is related to the edge of the Brillouin zone. In the case
of dilute quantum gases, a natural cut-off scale arises from the range $r_{0}$
of the interatomic potential: on scales smaller than this the interaction
amplitude cannot be represented any more by expression (\ref{Vk}) and should
be suppressed. The range of the interatomic potential for sodium is estimated
through the Van der Waals radius $r_{0}=2.4$ nm, related by $r_{0}=\frac{1}%
{2}(mC_{6}/\hbar^{2})^{1/4}$ to the Van der Waals coefficient $C_{6}=1556$
\cite{Derevianko}. In units of $\xi^{-1}$, this places the cut-off at
$K_{c}\approx200$ for the parameters listed in this section.

\subsection{Free energy and critical coupling strength}

\begin{figure}
[ptb]
\begin{center}
\includegraphics[
height=3.7334in,
width=4.8218in
]%
{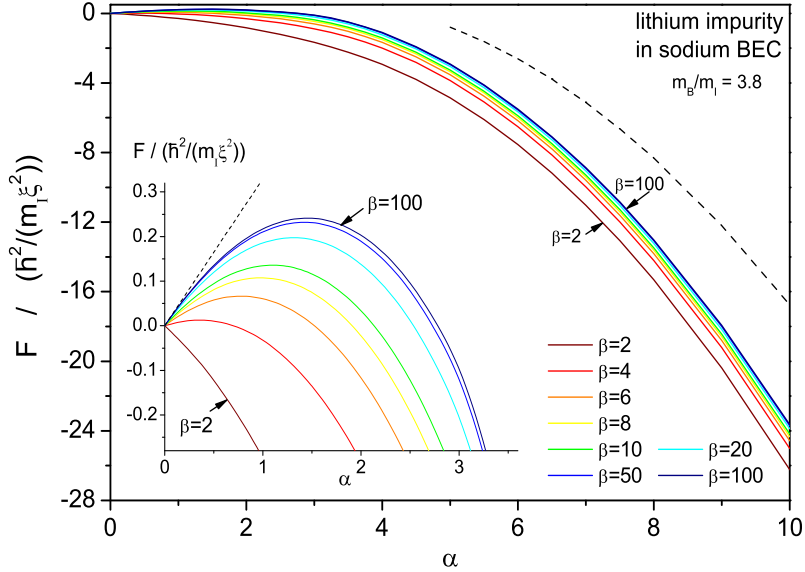}%
\caption{(Color online) The polaronic contribution (\ref{result}) to the free
energy of an impurity in a condensate is shown as a function of the coupling
constant $\alpha$, for different values of the temperature. From top to
bottom, these are $\beta=100,50,20,10,8,6,4,2$, corresponding in our example
to $T=3.5,7,18,35,44,58,88$ and $175$ nK, respectively. The dashed curve shows
the strong-coupling Pekar approximation. The inset zooms in on the small
$\alpha$ region: the dashed line in inset shows the second order perturbation
result. }%
\label{figenerg}%
\end{center}
\end{figure}

Figure \ref{figenerg} shows the results for the free energy as a function of
$\alpha$, using $m_{B}/m_{I}=3.8$ and a wave number cut-off at $K_{c}=200$,
for different values of $\beta$, ranging from $\beta=100$ to $\beta=2$, where
thermal depletion of the condensate starts to be appreciable.\ Standard
cooling schemes, as mentioned, can cool down to roughly $\beta=15$. The dashed
line shows the weak-coupling perturbative result valid for temperature zero
(and without any wave number cut-off). As predicted by Timmermans and
co-workers, at small $\alpha \ll1$ and $\beta \gg1$ the polaronic contribution
to the energy is positive. It reaches a maximum around $\alpha=1.5-2.0$ and
then decreases. The polaronic energy contribution becomes negative (indicative
of a transition from an unbound state to a self-trapped state) at a critical
coupling strength $\alpha_{c}\simeq3$ for $T\rightarrow0$. This critical value
goes to zero for increasing temperatures. The perturbative solution is seen to
fit well at low coupling. At larger coupling, the dashed curve indicates the
strong-coupling variational result (\ref{strong}): it is significantly larger
than the result of the full variation with $M$ and $\Omega$ free parameters,
indicating that a Gaussian wave function may not be as suitable as it is for
Fr\"{o}hlich polarons in the solid state. We emphasize that the polaronic
energy calculated here is the contribution from (\ref{Hpol}) and does not
contain the terms $E_{GP}+N_{0}V_{IB}(\mathbf{0})$ which have a known
dependence on the various tunable parameters $N_{0},a_{IB}$ and $a_{BB}$, and
which complicate the experimental determination of the polaronic energy
contribution. To observe polaronic effects, it may be more straightforward to
measure the shift in effective mass of the impurity.%

%TCIMACRO{\FRAME{ftbpFU}{4.8218in}{3.7334in}{0pt}{\Qcb{(Color online) The
%polaronic contribution (\ref{result}) to the free energy of an impurity in a
%condensate is shown as a function of the coupling constant $\alpha$, for
%different values of the temperature. From top to bottom, these are
%$\beta=100,50,20,10,8,6,4,2$, corresponding in our example to
%$T=3.5,7,18,35,44,58,88$ and $175$ nK, respectively. The dashed curve shows
%the strong-coupling Pekar approximation. The inset zooms in on the small
%$\alpha$ region: the dashed line in inset shows the second order perturbation
%result. }}{\Qlb{figenerg}}{figure1.eps}{\special{ language "Scientific Word";
%type "GRAPHIC";  maintain-aspect-ratio TRUE;  display "USEDEF";
%valid_file "F";  width 4.8218in;  height 3.7334in;  depth 0pt;
%original-width 4.1779in;  original-height 3.2283in;  cropleft "0";
%filename 'figure1.eps';file-properties "XNPEU";}}}%
%BeginExpansion
%EndExpansion

\subsection{Effective mass increase}

The effective polaron mass $m_{pol}$ can be derived from the path-integral
propagation of a particle from $r(0)$ to a nearby point $r(\tau)=r(0)+v\tau$
by casting the resulting transition amplitude in the form $\exp \{-m_{pol}%
v^{2}\tau/2\}.$ Feynman notes that this procedure gives a value for $m_{pol}$
which is always within a few percent of $1+M$ with $M$ the variationally
optimal mass $M$ of the trial model\cite{Feynmanbook,Schultz}. Figure 2 shows
the result for $M$ as a function of $\alpha$, for different values of the
temperature. For small values of the coupling strength, the mass increases
linearly with $\alpha$ as predicted by perturbation theory. However, near
$\alpha \approx3.5$, the behavior changes and the mass increases rapidly.%

%TCIMACRO{\FRAME{ftbpFU}{4.7952in}{4.0122in}{0pt}{\Qcb{(Color online) The
%variational mass parameter $M$ is shown as a function of $\alpha.$ The
%effective polaron mass is to a good approximation given by $m_{pol}%
%=m_{I}(1+M/m_{I}).$ The curves correspond to different temperatures
%($\beta=100,50,20,10,8,6,4,2$, from top to bottom), at cut-off $K_{c}=200.$
%The inset shows the effect of cutoff ($K_{c}=100,200,400,1000$, from top to
%bottom), at $\beta=10$.}}{\Qlb{fig2mass}}{figure2.eps}%
%width 4.7952in;  height 4.0122in;  depth 0pt;  original-width 3.8303in;
%original-height 3.2007in;  cropleft "0";  croptop "1";  cropright "1";
%cropbottom "0";  filename 'figure2.eps';file-properties "XNPEU";}}}%
%BeginExpansion
\begin{figure}
[ptb]
\begin{center}
\includegraphics[
height=4.0122in,
width=4.7952in
]%
{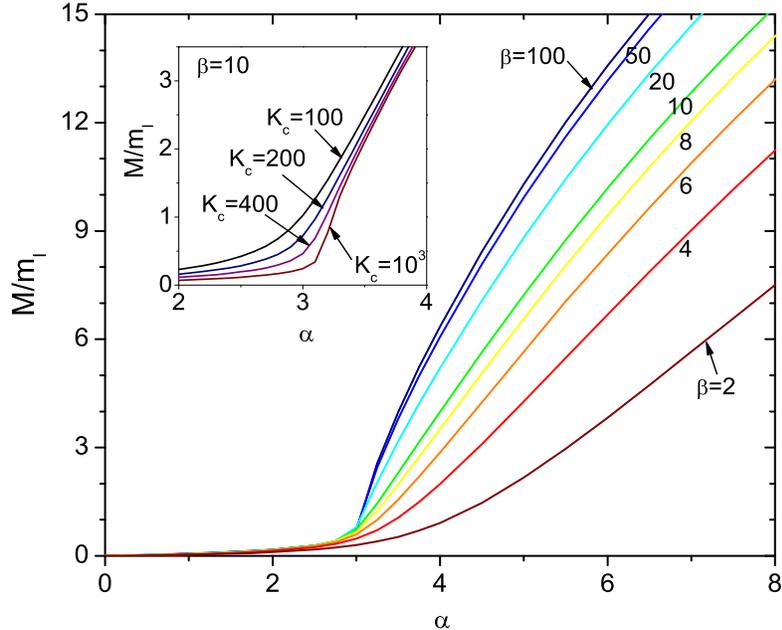}%
\caption{(Color online) The variational mass parameter $M$ is shown as a
function of $\alpha.$ The effective polaron mass is to a good approximation
given by $m_{pol}=m_{I}(1+M/m_{I}).$ The curves correspond to different
temperatures ($\beta=100,50,20,10,8,6,4,2$, from top to bottom), at cut-off
$K_{c}=200.$ The inset shows the effect of cutoff ($K_{c}=100,200,400,1000$,
from top to bottom), at $\beta=10$.}%
\label{fig2mass}%
\end{center}
\end{figure}
%EndExpansion

The low-energy Bogoliubov excitations are sound waves, with a dispersion
qualitatively similar to acoustic phonons. The effective mass of acoustic
polarons -- electrons interacting with acoustic phonons -- shows a jump of
several orders of magnitude in the effective mass (cf. \cite{acoustipol})
above a critical coupling strength. In the present case we also note a faster
increase in the mass above a critical coupling strength, even though the
interaction amplitude is different from that between acoustic phonons and
electrons. In the dilute atomic gas, the transition is much less dramatic, and
becomes smoother as temperature increases. We believe the smoothness of the
crossover is not an artifact of the path-integral formalism, since in the case
of the acoustic polarons it is the same formalism that predicts a
discontinuous jump. For acoustic polarons, it is also predicted that the
sharpness of the transition depends strongly on the cutoff: above a critical
value of $K_{c}$ a discontinuity appears in the mass as a function of $\alpha
$. So it is worthwhile to study the dependence of $M$ on the value of a
cut-off to the $k$ integrations in (\ref{result}) also in the present case. We
find that increasing the cutoff sharpens the transition (as can be seen in the
inset of Fig. \ref{fig2mass}), but no discontinuity arises as it does for the
acoustic polaron. The difference is due to the fact that although the
Bogoliubov excitation dispersion and the interaction amplitude show a similar
$k$-dependence as in the case of acoustic polarons, this is only true in the
limit $k\ll1,$ and large deviations already occur for $k\approx1$, the
relevant length scale of the problem. Yet even though the transition is not as
abrupt as for acoustic polarons, it is possible to distinguish two regimes. In
conjunction with the crossover from positive to negative values of the free
energy, this is again indicative of a transition between an quasi-free
(unbound) impurity and self-trapping for the impurity.

\subsection{Polaron radius}

The second variational parameter, $\Omega$, is linked to the polaron radius.
Within the model system described by the action functional $\mathcal{S}_{0}$,
expression (\ref{model}), the expectation value of the square of the relative
coordinate for the impurity -- boson mass system is given by%
\begin{equation}
\left \langle r^{2}\right \rangle =\frac{3}{2\Omega}\frac{M}{1+M}\coth \left(
\frac{\beta \Omega}{2}\right)
\end{equation}
The square root of this is a measure of the localization length of the
impurity wave function. For strong coupling, this expectation value converges
to the expectation value with respect to the variational wave function
formulated by Landau and Pekar\cite{Pekar1946}.%

%TCIMACRO{\FRAME{ftbpFU}{4.6134in}{3.4941in}{0pt}{\Qcb{(Color online) The
%polaron radius in units of the healing length is plotted as a function of the
%coupling constant $\alpha$, for different temperatures. From top to bottom,
%$\beta=100,50,20,10,8,6,4,2$. Around $\alpha\approx3$ nonmonotonous behavior
%develops as a function of $\alpha$ when the temperature is reduced.}%
%}{\Qlb{fig3size}}{figure3.eps}{\special{ language "Scientific Word";
%type "GRAPHIC";  maintain-aspect-ratio TRUE;  display "USEDEF";
%valid_file "F";  width 4.6134in;  height 3.4941in;  depth 0pt;
%original-width 3.9401in;  original-height 2.9776in;  cropleft "0";
%filename 'figure3.eps';file-properties "XNPEU";}}}%
%BeginExpansion
\begin{figure}
[ptb]
\begin{center}
\includegraphics[
height=3.4941in,
width=4.6134in
]%
{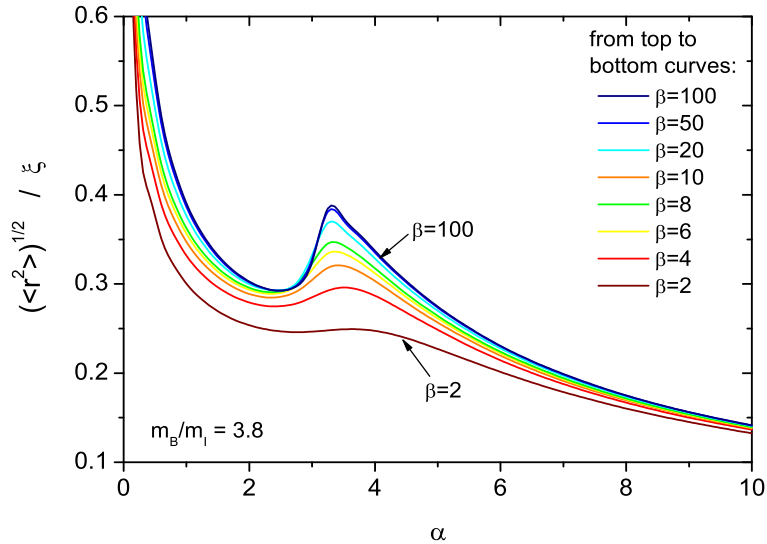}%
\caption{(Color online) The polaron radius in units of the healing length is
plotted as a function of the coupling constant $\alpha$, for different
temperatures. From top to bottom, $\beta=100,50,20,10,8,6,4,2$. Around
$\alpha \approx3$ nonmonotonous behavior develops as a function of $\alpha$
when the temperature is reduced.}%
\label{fig3size}%
\end{center}
\end{figure}
%EndExpansion
Fig. (\ref{fig3size}) shows the polaron radius $\sqrt{\left \langle
r^{2}\right \rangle }$ as a function of the coupling strength, for different
values of $\beta,$ at a cut-off $K_{c}=200$. As the coupling strength is
increased, the polaron radius decreases, indicating a stronger confinement of
the impurity wave function. In particular the large $\alpha$ behavior might
still depend on the cut-off, although a $1/\alpha$ dependence is expected
independently of $K_{c}$. Near $\alpha \approx3-3.5,$ we find that a local
maximum develops and the behavior of the polaron size as a function of
$\alpha$ becomes non-monotonic. It is not clear whether this local maximum is
an intrinsic feature of the BEC-impurity polaron, or whether it is an artifact
from the particular variational model used.

\section{Conclusion}

When the Bogoliubov approximation applies, the Hamiltonian describing the
impurity in a condensate can be cast in the form of the\ Fr\"{o}hlich polaron
Hamiltonian. The physics becomes similar to that of a Fr\"{o}hlich polaron,
where for the impurity in the BEC the Bogoliubov excitations have taken the
role of the phonons and the interaction strength is related to the
impurity-boson and boson-boson scattering lengths. The most accurate
description of polaron physics in the case of electron-phonon interactions is
given by Feynman's variational treatment, which moreover allows to study the
temperature dependence of observables such as the effective mass and the free
energy. We have applied a Jensen-Feynman path-integral type calculation to the
case of the impurity in a condensate and derived expression (\ref{result}) for
the free energy. Both in the free energy and the effective mass, a critical
value of the coupling strength $\alpha \approx3.5$ can be identified where the
system crosses over from the weak-coupling to the strong-coupling regime.
These regimes show a qualitatively different behavior of the effective mass
and free energy. The sharp increase in the the effective mass in the
strong-coupling regime hints at a transition from an quasi-free state to a
large-mass state similar to that for acoustic polarons. It has been pointed
out\cite{acoustipol} that a transition from the quasi-free regime to the
large-mass regime is impossible to realize in most semiconductors and III-V
compounds, even in alkali halides. However, the present results indicate that
it might be attainable in ultracold gases. For this purpose, we investigated
the experimentally relevant values of the system parameters, and derived a
useful expression (\ref{alpha}) relating $\alpha$ to the various parameters in
the case of a condensate in the Thomas-Fermi regime. This opens up the
prospect to reach and investigate the strong-coupling regime in ultracold
gases, whereas this regime has hitherto be inaccessible in the solid state.

\begin{acknowledgments}
This work is supported by FWO-V project G.0180.09N, and projects G.0115.06,
G.0356.06, G.0370.09N, the WOG WO.033.09N (Belgium), and INTAS Project no.
05-104-7656. J.T. gratefully acknowledges support of the Special Research Fund
of the University of Antwerp, BOF NOI UA 2004. M.O. acknowledges financial
support by the ExtreMe Matter Institute EMMI in the framework of the Helmholtz
Alliance HA216/EMMI.
\end{acknowledgments}

\bigskip

\end{document}